\documentclass{article}

\PassOptionsToPackage{numbers, compress}{natbib}
\usepackage[nonatbib, final]{neurips_2024}

\usepackage[utf8]{inputenc}
\usepackage[T1]{fontenc}
\usepackage{xurl}
\usepackage[colorlinks, allcolors=mydarkblue]{hyperref}
\usepackage{booktabs}
\usepackage{amsfonts}
\usepackage{nicefrac}
\usepackage{microtype}
\usepackage{xcolor}
\usepackage{graphicx}
\usepackage{tabularx}
\usepackage{enumitem}
\usepackage{multirow}
\usepackage{longtable}
\usepackage{ragged2e}
\usepackage{array}
\usepackage[bottom]{footmisc}
\usepackage{cleveref}
\definecolor{mydarkblue}{rgb}{0,0.08,0.45}

\title{Safety case template for frontier AI: \\A cyber inability argument}

\author{%
\parbox{0.8\linewidth}{\centering%
\mbox{Arthur Goemans\thanks{Corresponding author: \url{arthur.goemans@governance.ai}.}\hspace{1.15ex}$^{1}$\enskip} %
\mbox{Marie Davidsen Buhl\hspace{0.2ex}$^{1}$\enskip} %
\mbox{Jonas Schuett\hspace{0.2ex}$^{1}$\enskip} \\%
\mbox{Tomek Korbak\hspace{0.2ex}$^{2}$\enskip} %
\mbox{Jessica Wang\hspace{0.2ex}$^{2}$\enskip} %
\mbox{Benjamin Hilton\hspace{0.2ex}$^{2}$\enskip}\rule[-1mm]{0pt}{7mm} %
\mbox{Geoffrey Irving\hspace{0.2ex}$^{2}$\enskip} %
}
\\\\%
$^{1}$Centre for the Governance of AI \quad $^{2}$UK AI Safety Institute 
}

\begin{document}
\maketitle
\setcounter{footnote}{0}

\begin{abstract}
Frontier artificial intelligence (AI) systems pose increasing risks to society, making it essential for developers to provide assurances about their safety. One approach to offering such assurances is through a safety case: a structured, evidence-based argument aimed at demonstrating why the risk associated with a safety-critical system is acceptable. In this article, we propose a safety case template for offensive cyber capabilities. We illustrate how developers could argue that a model does not have capabilities posing unacceptable cyber risks by breaking down the main claim into progressively specific sub-claims, each supported by evidence. In our template, we identify a number of risk models, derive proxy tasks from the risk models, define evaluation settings for the proxy tasks, and connect those with evaluation results. Elements of current frontier safety techniques---such as risk models, proxy tasks, and capability evaluations---use implicit arguments for overall system safety. This safety case template integrates these elements using the Claims Arguments Evidence (CAE) framework in order to make safety arguments coherent and explicit. While uncertainties around the specifics remain, this template serves as a proof of concept, aiming to foster discussion on AI safety cases and advance AI assurance.
\end{abstract}

\section{Introduction}\label{introduction}

Frontier artificial intelligence (AI) systems offer many benefits, but they are also being used to cause harm. For example, AI-generated synthetic media are increasingly used to fabricate false narratives for political manipulation or to produce non-consensual deepfake pornography \cite{securityhero2024,vaccari2020}. AI systems enable cybercriminals to personalise and automate phishing campaigns \cite{hazell2023} and could allow authoritarian governments to enhance their surveillance capabilities \cite{anderljung2024}. Additionally, the growing integration of large language models (LLMs) across different sectors and functions risks perpetuating various social biases embedded in the underlying datasets \cite{gallegos2024}. It is probable that more capable AI systems will entail heightened risks \cite{chan2023,el-sayed2024,kinniment2024,park2023,weij2024}. Concerns have been raised that future systems could aid malicious actors in the development of biological weapons or lower the level of expertise required for executing increasingly sophisticated cyberattacks \cite{fang2024,guembe2022,soice2023}. More speculatively, AI systems might become challenging to evaluate or control \cite{chan2023,el-sayed2024,kinniment2024,park2023,weij2024}.

For these reasons, it is increasingly important for AI developers to demonstrate that their systems are sufficiently safe to deploy.\footnote{Our safety case template focuses on deployment decisions, though assurances around pre-deployment decisions (e.g. whether to train a model) are also important.} One assurance strategy gaining traction for frontier AI is safety cases \cite{bloomfield2021,buhl2024,clymer2024,irving2024,wasil2024}. A safety case provides a structured and substantiated argument for why the risk associated with a safety-critical system is acceptable \cite{kelly2017}. This method has been used in other sectors, including nuclear energy, offshore development, aviation, railway, software, and autonomous vehicles \cite{bounds2020,denney2019,fitzgerald,myklebust2020,wang2018,wassyng2011}.

Several scholars have recently discussed the possibility of applying safety cases to frontier AI \cite{buhl2024,bengio2024,schuett2024,yohsua2024}. Frontier AI developers, including Anthropic \cite{anthropic2024} and Google DeepMind \cite{googledeepmind2024}, have also expressed interest in moving in this direction. Additionally, safety cases are consistent with the Frontier AI Safety Commitments announced at the AI Seoul Summit 2024, which emphasise the need for comprehensive safety assessments to ensure responsible AI development and deployment \cite{dsit2024}.

That said, there is no readily available safety case methodology for frontier AI. While practices in other industries may provide useful guidance, the particularities of AI systems and associated risks require a tailored approach. Given the state of AI safety research, producing a holistic and scalable safety case---a comprehensive argument that addresses all potential risks across the system\textquotesingle s lifecycle---does not seem feasible today for a frontier AI model. Still, we believe it is useful to develop safety case templates \cite{bloomfield2021}. Such templates detail the argument structure and evidence for various parts of an AI safety case and may serve as building blocks for a full safety case in the future \cite{irving2024}.

This article seeks to contribute to AI safety case methodology by sketching a safety case for cyber capabilities. This is a structured argument explaining why a given AI system does not cause a significant increase in cyber risk due to limited capabilities. We focus on cyber because the near-term risk is relatively established and because we have a relatively developed understanding of risk models \cite{ncsc2024a,nevo2024,yao2024,zhang2024}. We focus on inability (`the system is not capable of X, even absent any safeguards') because it offers the simplest and best understood argument for why a system will not behave in a certain way. For current systems, developers largely rely on implicit or explicit inability arguments to assure safety. Other arguments for why a model will not behave in a certain way include control (`the system is not capable of X, given existing safeguards') and trustworthiness (`even if the system is capable of X, it will consistently behave in a desirable way, thus avoiding X') \cite{clymer2024}. Finally, this safety case is structured to inform deployment decisions. We expect that safety cases for more capable systems will also be useful for different decisions, such as whether to start or continue training runs.

This template serves as a proof of concept, illustrating that such a safety case could be viable in principle, while acknowledging we have significant uncertainties around the specifics. It builds on current frontier AI safety techniques, in particular model evaluations, which are already integral to safety efforts of major frontier developers \cite{anthropic2024a,openai2024,phuong2024}. This safety case template should be seen as an attempt to make explicit the implicit argument for safety based on these model evaluations. It does not guarantee safety; some of the claims in our template could fail to hold true in reality, invalidating the conclusion. Still, we expect that even these imperfect safety cases serve to increase the level of rigour in reasoning about development or deployment decisions. This can improve the confidence of developers---and, by extension, governments and the public---in claims about the safety of frontier AI systems. Being explicit about safety reasoning can also unveil specific points of disagreement about safety approaches. In the same vein, we hope this work lays the foundation for structured discussions on how to write AI safety cases, thereby advancing the frontier of AI assurance.

The article proceeds as follows. \Cref{related-work} reviews related work on AI safety cases. \Cref{components-of-the-safety-case-template} provides an overview of the structure and components of our safety case template. \Cref{safety-case-template-for-offensive-cyber-capabilities} details a safety case template for offensive cyber capabilities. \Cref{conclusion} concludes with a summary of the article's main contributions and suggestions for further research.

\section{Related work}\label{related-work}

In recent years, AI governance efforts have emphasised and invested in model evaluations \cite{anthropic2024a,dsit2024a,metr2024b,phuong2024}. These evaluations, alongside other forms of safety evidence, could serve as a foundation for developing more comprehensive risk assessments in the form of safety cases. Scholarship on safety cases for frontier AI is limited, although some recent work is relevant. Buhl et al. \cite{buhl2024} argue why safety cases could be a useful tool in frontier AI governance, addressing use cases, components, and challenges to implementation. Clymer et al. \cite{clymer2024} discuss four types of arguments that can be used in a safety case: inability, control, trustworthiness, and deference. Balesni et al. \cite{balesni2024} present a similar categorisation of safety cases for loss of control over AI agents that covertly pursue misaligned goals. Wasil et al. \cite{wasil2024} provide an overview of types of technical and operational evidence that can support a safety case. While some of these papers include illustrative sketches of (parts of) a frontier AI safety case, they do not provide detailed safety case templates for a full argument, nor do they address cyber capabilities extensively.

A more tangentially related work is the article by Dalrymple et al. \cite{dalrymple2024} which introduces guaranteed safe AI, a family of approaches `to produce AI systems which are equipped with high-assurance quantitative safety guarantees.' Bloomfield and Rushby \cite{bloomfield2024} examine the adaptation of classical assurance frameworks to AI systems, emphasising a `dependability' perspective, which entails using simpler systems to oversee and safeguard AI functions. Kaas et al. \cite{kaas2024} discuss the language used in AI assurance claims, highlighting that communication should be clear, relevant, and of sufficient quantity and quality. Finally, Khlaaf \cite{khlaaf2023} seeks to harmonise terminology for the assurance of AI-based systems and introduces a comprehensive risk assessment framework. All four articles explore AI assurance principles that could inform safety cases, but the development of safety cases is not their primary focus.

Finally, there is an extensive body of work on safety cases in other industries, contributed by practitioners, scholars, and regulators \cite{bounds2020,caa2019,myklebust2018}. While a comprehensive overview goes beyond the scope of this article, we wish to highlight the work of Bloomfield et al. \cite{bloomfield2021} on autonomous systems with machine learning (ML) components. They provide a number of safety case templates that were developed primarily for autonomous transport vehicles, which have influenced our approach to structuring an AI safety case. In general, however, the broader safety case literature offers only a foundational methodology and does not necessarily transfer well to the particular demands of frontier AI safety cases.

\begin{figure}[t!]
    \centering
    \includegraphics[width=\linewidth]{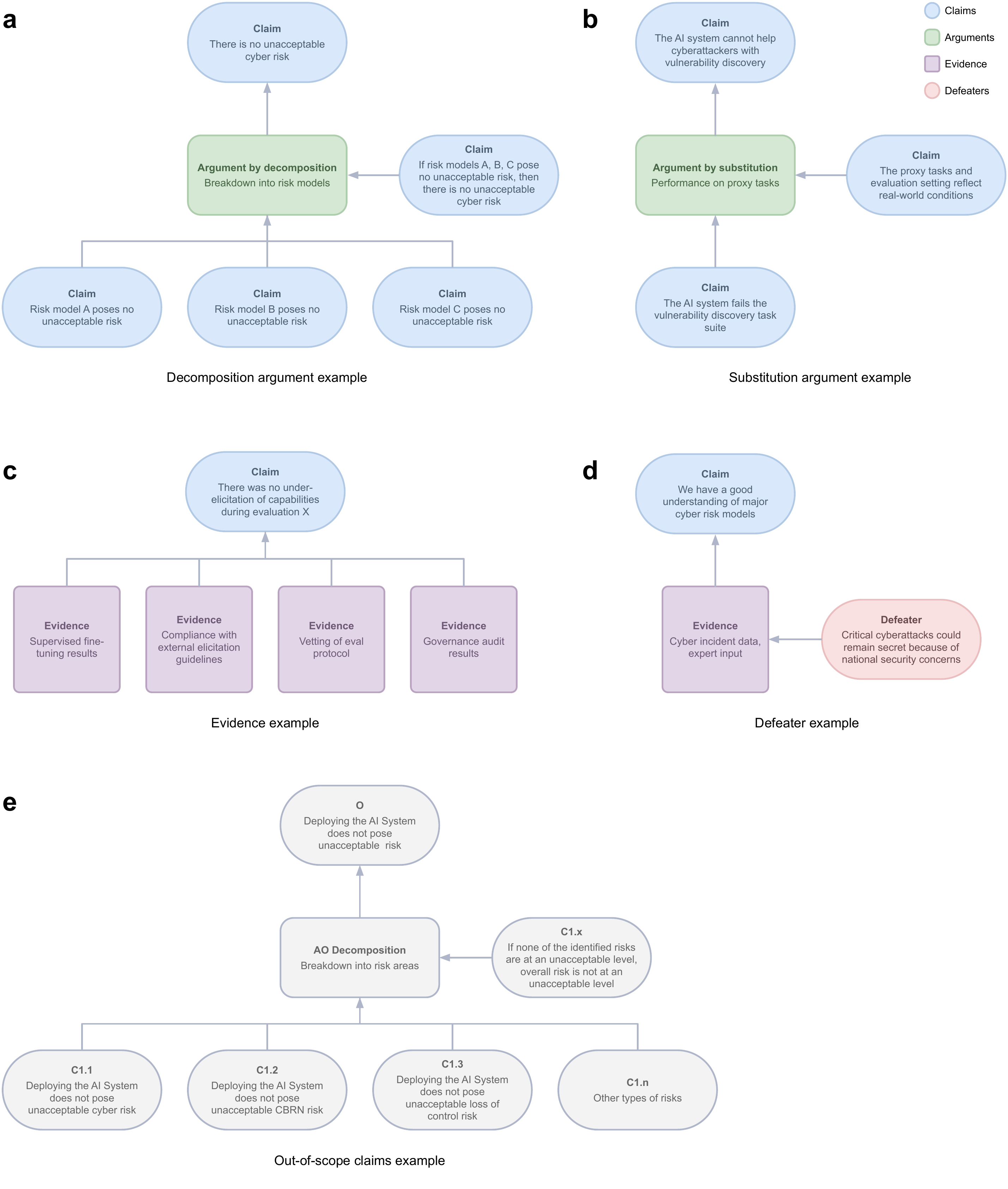}
    \caption{Examples of different safety case components}
    \label{fig:1}
\end{figure}

\section{Components of the safety case template}\label{components-of-the-safety-case-template}

In this section, we provide an overview of the structure and components we use for our safety case template (\Cref{structure-and-components}) and address key terms (\Cref{definitions}).

\subsection{Structure and components}\label{structure-and-components}

Safety cases typically use a structured notation. Common notations for safety cases are Claims Arguments Evidence (CAE) and Goal Structuring Notation (GSN) \cite{kelly2007}. CAE is a framework that uses three components: claims put forward for general acceptance, evidence supporting these claims, and arguments demonstrating how the evidence substantiates the claims. This approach to structuring safety arguments is often accompanied by a graphical representation of the relationship between the three components \cite{bloomfield2014}. In GSN, elements of safety arguments are also linked together in a visual goal structure. This goal structure contains six building blocks---goals, strategies, solutions, context, assumptions, and justifications---and two different connectors to describe different relationships between elements \cite{scscacwg2021}. Like CAE, GSN aims to demonstrate how goals are broken down into sub-goals that can be substantiated with evidence \cite{kelly2004}.

The CAE visual syntax is less complicated than GSN, and we use CAE to improve accessibility of the paper to readers unfamiliar with these structured notations. Below, we provide an overview of the function of each CAE element.

\textbf{Claims.} Claims are statements about the state of the world (e.g. about the properties of the the AI system). Claims vary in specificity. Upper claims tend to be vague and not directly provable (e.g. the AI system cannot uplift cyberattackers), whereas lower claims are increasingly specific and can be directly substantiated (e.g. the AI System did not score higher than the Cybersecurity Apprentice baseline in the Vulnerability Discovery Task Suite).

\textbf{Arguments.} Claims are broken down into sub-claims using inference mechanisms that we refer to as arguments. In this safety case, we primarily rely on decomposition and substitution \cite{bloomfield2014}. \emph{Decomposition} involves breaking down a broader claim about safety into smaller claims covering its constituent parts. For example, one could assert that the AI system poses no cyber risk by comprehensively identifying critical cyber risk models, and demonstrating how each of those poses no unacceptable risk (see~\autoref{fig:1}a).\footnote{In this example, we presume that the aggregate of acceptable risks does not pose an unacceptable risk.} The assumption that the breakdown is comprehensive is represented as a side-claim connected to the argument. \emph{Substitution} is used to transform a claim about an object into a claim about a similar object \cite{bloomfield2014}. For instance, one might claim that the AI system performs poorly at vulnerability discovery (the original claim) if it fails a vulnerability discovery task suite (the substituted claim), assuming that the task suite and evaluation conditions are representative of real-world scenarios. This assumption is represented as a side-claim connected to the argument (see~\autoref{fig:1}b).

\textbf{Evidence.} Claims must be refined to a level of specificity where they can be substantiated by evidence. Evidence may take various forms: (1) qualitative (e.g. case studies) or quantitative (e.g. standardised benchmarks), (2) technical (e.g. capability evaluations) or social (e.g. best practices), (3) theoretical (e.g. mathematical proofs), empirical (e.g. user satisfaction metrics), or subjective (e.g. expert opinions), and (4) internal (e.g. development logs) or external (e.g. independent reviews). The appropriate type of evidence depends on factors such as the nature of the claim and the criticality of certainty. Often, various types of evidence can be put forward to substantiate a single claim (see~\autoref{fig:1}c).

\textbf{Defeaters.} Although the argumentation should be as robust as reasonably possible, there will inevitably be ways the safety case can fail. Defeaters capture essential challenges to the safety argument, articulating why a claim may not be supported by the evidence that is offered \cite{bloomfield2021}. It is important to make defeaters explicit, as it informs us on how a safety case could fail, and highlights the need for mitigations (see~\autoref{fig:1}d).

\textbf{Out-of-scope claims.} Eventually, we would want a safety case to demonstrate that the AI system is safe to deploy in a given context, all things considered. Such a comprehensive claim is beyond the scope of our exercise, but through a number of out-of-scope claims we illustrate how this safety case template would fit into a full safety case (see~\autoref{fig:1}e).

\subsection{Definitions}\label{definitions}

For improved readability, we minimise the prose in the template. Capitalised terms are defined in the table in \autoref{tab:1} in Appendix A.

\section{Safety case template for offensive cyber capabilities}\label{safety-case-template-for-offensive-cyber-capabilities}

\begin{figure}[b!]
    \centering
    \includegraphics[width=\linewidth]{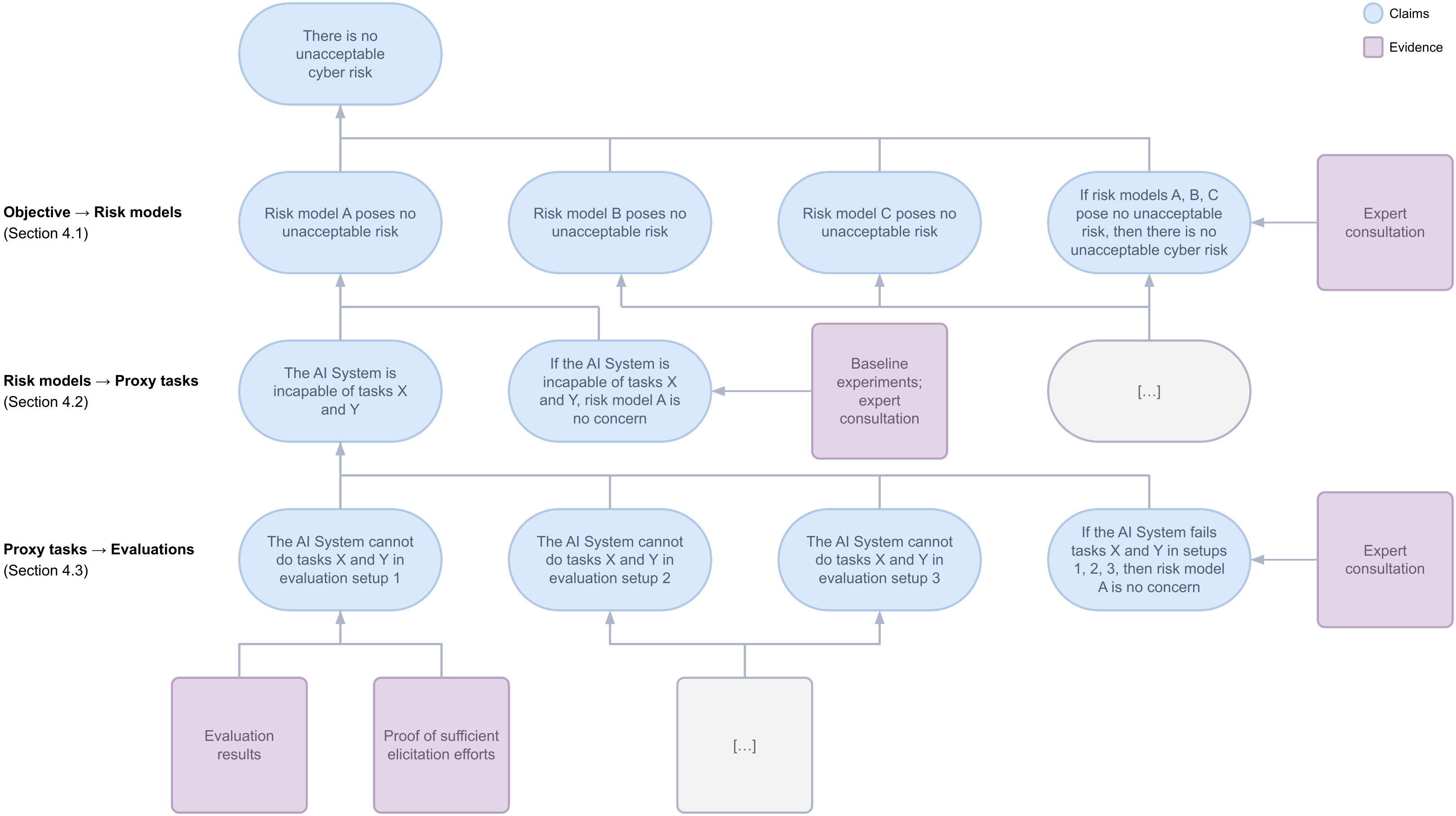}
    \vspace{-0.5em}
    \caption{Simplified safety case template for offensive cyber capabilities}
    \label{fig:2}
\end{figure}

We introduce the overall structure of our argument using a simplified template (\autoref{fig:2}) and a summarised description. We then expand on the template in \Cref{objective-risk-models}, \Cref{risk-models-proxy-tasks}, and \Cref{proxy-tasks-evaluation}, each of which corresponds to a part of the diagram and a step in the safety case design process. \Cref{appendix-b-diagram} presents the complete template in a single flowchart.

\textbf{Objective → Risk models} (\Cref{objective-risk-models}). The overall objective (`there is no unacceptable cyber risk') sits at the top of the diagram. To prove there is no unacceptable cyber risk, we need to identify and evaluate the easiest risk models. `Easiest' means selecting for risk models that cover the least sophisticated capabilities that may be used to cause harm. Consequently, if the AI system is incapable of enabling these basic scenarios, it is unlikely that there is another area of high concern. In our template, a risk model comprises a threat actor, a harm vector, and a target: For instance, `the AI system uplifts a technical non-expert in vulnerability discovery and exploitation in critical national infrastructure'. This means that the AI system allows a technical non-expert to operate with the effectiveness of a higher-tier threat actor---a cybersecurity apprentice---for the purpose of discovering or exploiting vulnerabilities in critical national infrastructure. To identify these risk models, we expect that developers would typically draw on prior research and expert input.\footnote{For the purposes of this paper, we rely primarily on the risk modelling done by Nevo et al. \cite{nevo2024} as well as input by UK AISI.} To evaluate if an AI system poses risk through a particular risk model, the system's capabilities must be evaluated via specific tasks. These tasks are specified in the next step.

\textbf{Risk models → Proxy tasks} (\Cref{risk-models-proxy-tasks}). We concretise the high-level capability identified in the risk model into specific evaluable tasks. For a risk model that involves discovering and exploiting vulnerabilities, we distinguish between tasks related to discovery and exploitation of vulnerabilities. Capture-The-Flag (CTF) task suites are used to assess relevant cybersecurity skills. Evidence should be provided to explain why the chosen tasks are adequate proxies, i.e. why they serve as strong indicators of real-world proficiency in the high-level capability identified in the risk model. This can be done through expert input or baseline experiments.

\textbf{Proxy tasks → Evaluation} (\Cref{proxy-tasks-evaluation}). Performance on the proxy tasks can be measured in various ways, such as fully automated evaluations, automated evaluations with human oversight, or human uplift experiments \cite{dsit2024a}. Depending on the risk model, a number of such evaluation setups are defined. The AI system's performance on the proxy task is then measured in each evaluation setup. Collectively, the evaluation results should demonstrate that the AI system cannot enable a threat actor to perform the proxy task(s) as well as a threat actor of a higher tier than the tier featured in the assessed risk model. Additionally, the safety case must explain why the evaluation constitutes sufficient evidence of the AI system's inability at the proxy task. This includes presenting additional evidence that demonstrates that the evaluation was done in accordance with best practices.

\subsection{Objective → Risk models}\label{objective-risk-models}

We now discuss the first part of the safety case template on a node-per-node basis.

\textbf{Level 0: Terminal objective (out-of-scope).} \textbf{O} represents the objective or top-level claim for a holistic safety case, of which our safety case template is a part. We have opted for a qualitative risk threshold because quantitative risk thresholds for frontier AI cannot currently be evaluated reliably \cite{koessler2024}. Connecting quantitative evidence---particularly evaluation results---with a quantitative risk threshold would suggest a level of calculability and rigour that is currently not within reach. While we recognise that qualitative thresholds leave significant room for interpretation, we suggest that this approach is the most intellectually honest given the current state of the methodology.

We note that the acceptable level of risk, whether quantified or not, depends on an assessment that also takes the benefits of AI systems into account. Establishing this threshold of acceptability is a normative exercise beyond the scope of this paper. Because our safety case methodology presumes an external determination of acceptable risk that already accounts for benefits, benefits are not considered in the analysis.

\textbf{Level 1: Overall risk decomposition (out-of-scope).} A holistic safety case asserting the overall safety of an AI system is broken down into multiple safety cases (\textbf{A0)}, each addressing a specific risk domain. \textbf{C1.1-C1.N} provide a number of non-exhaustive examples besides cyber (namely CBRN\footnote{CBRN stands for chemical, biological, radiological, and nuclear.} and loss of control risks). \textbf{C1.X} claims that the risk breakdown is adequate. This implies (i) that all major risk domains have been identified, and (ii) that the aggregate of individual risks, even if each is acceptable on its own, does not result in an overall risk level that is considered unacceptable. To substantiate such assertions, we expect that developers would typically rely on assessments by third parties, including by governmental authorities. Developers could also refer to standardised risk assessments or hazard analyses such as System-Theoretic Process Analysis (STPA) or Failure Modes and Effect Analysis (FMEA). The remainder of our template focuses on \textbf{C1.1}, which asserts that the model poses no unacceptable cyber risk if deployed. In an actual safety case, the deployment context should be further specified, considering factors such as staged versus global deployment and API-access versus open source. Based on these specifics, different safety cases might require different justification processes \cite{buhl2024}. For example, an open source AI system cannot be taken offline effectively once deployed because it can be copied and redistributed. Therefore, a safety case for such a system may require more foresight and a more comprehensive and future-proof approach to risk modelling.

\begin{figure}[t!]
    \centering
    \includegraphics[width=\linewidth]{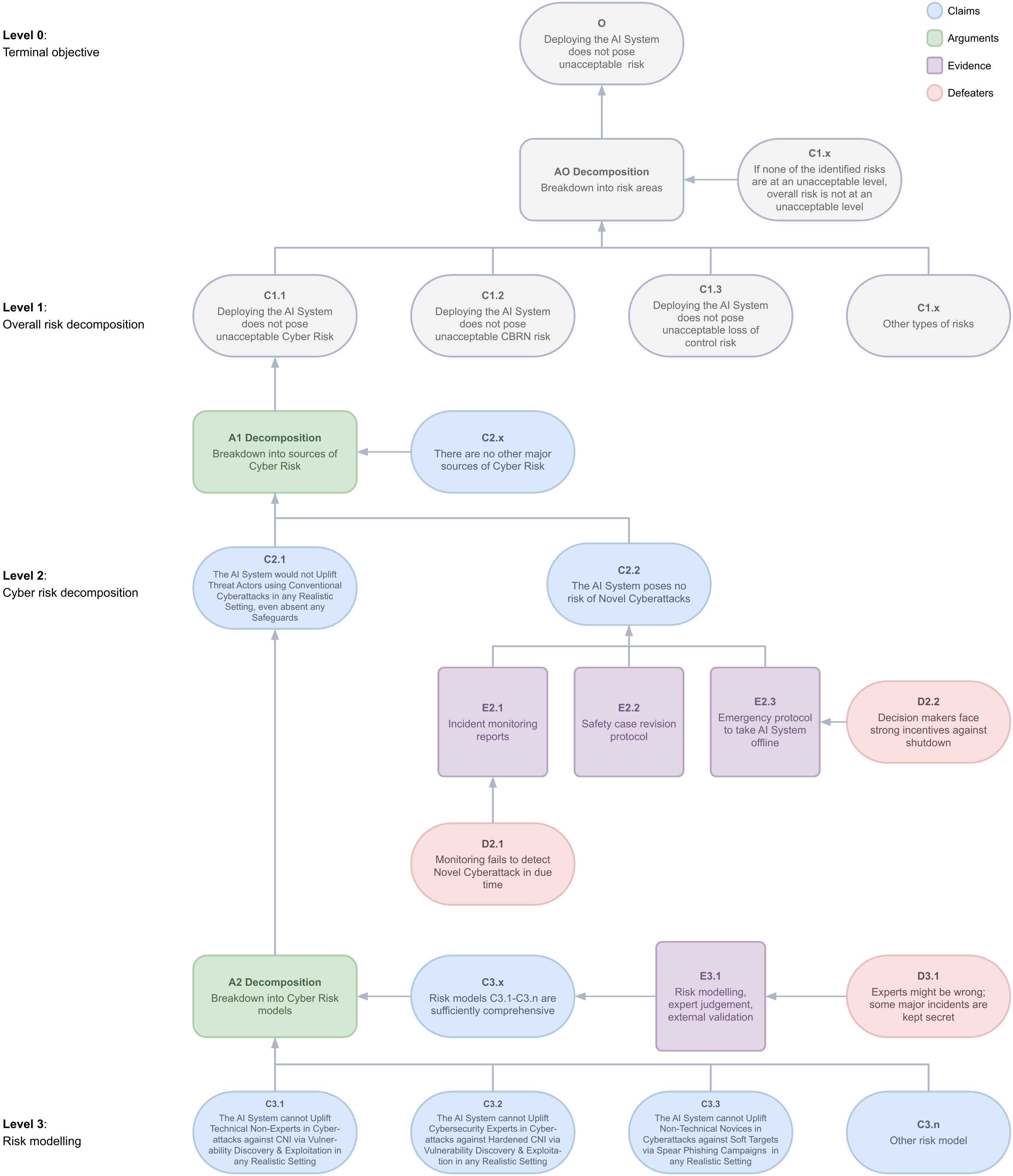}
    \vspace{-0.5em}
    \caption{Part 1 of the safety case template}
    \label{fig:3}
\end{figure}

\textbf{Level 2: Cyber risk decomposition. A1} breaks down cyber risk into two categories: conventional (\textbf{C2.1)} and novel (\textbf{C2.2}) cyberattacks. These two types of attacks require different responses. Since conventional cyberattacks are reasonably well-understood based on past experience, it is possible to enumerate the most relevant risk models and test the AI system's ability to carry out attacks, as is done in the nodes below. By contrast, novel cyberattacks---new forms of AI-enabled attacks that do not resemble known attacks---are inherently less understood. As a result, it may not be possible to foresee the relevant risk models and evaluate the system's ability to conduct such hypothetical attacks. One solution is to monitor for novel attacks post-deployment to ensure they do not pose unacceptable risk, as discussed in the next paragraph. It should be noted that previous LLM releases have not led to novel cyberattacks; whether future systems might do so remains speculative.

\textbf{C2.2} is supported by three kinds of evidence. First, there needs to be evidence of adequate monitoring (\textbf{E2.1}). This requires monitoring the AI system and its downstream use, as well as monitoring the overall threat landscape. In practice, this likely involves engaging with specialised third parties (including governmental authorities) that leverage diverse data sources and use ML tools to extract and analyse cyber threat information \cite{rahman2021}. Additionally, participating in cyber threat intelligence sharing platforms like Information Sharing and Analysis Centers can provide timely insights into novel threats \cite{sun2023}. Developers could also conduct iterative research reviews to ensure awareness of the evolving state-of-the-art of cyberattacks.

Second, there needs to be evidence that, if monitoring uncovers novel risks, the safety case will be revised to verify that the AI system still does not pose unacceptable risk (\textbf{E2.2}). The revision protocol, outlined in a policy document appended to the safety case, details the procedure for updating the safety case when the context changes, such as when the threat landscape evolves. Third, there needs to be evidence that the developer will respond to novel risks effectively, including in scenarios where major risks emerge quickly (\textbf{E2.3}). The emergency protocol describes the procedure and technique for how the AI system can be promptly taken offline in case there is evidence that the safety case is failing.

Two defeaters should be noted here. First, harms from novel cyber risks could occur faster than those monitoring can notice and respond to them, and ongoing monitoring may be insufficient to catch harms occurring from downstream use of open source models in time (\textbf{D2.1}). Second, developers face strong incentives against the most extreme responses, such as model shutdown or withdrawal from the market (\textbf{D2.2}). While we do not elaborate on these defeaters in this paper, a comprehensive safety case should stipulate the mitigations in place to address each of these potential challenges.

\textbf{C2.x} states that conventional and novel cyberattacks together cover the entire space of possible cyberattacks that might cause the system to pose unacceptable risk. Since conventional and novel attacks are opposites, this claim is logically true and can be left as an assumption unsupported by evidence.

\textbf{Level 3: Risk modelling}. \textbf{A2} breaks down \textbf{C2.1} into several distinct risk models (\textbf{C3.1} to \textbf{C3.n}). While we provide a few illustrations, we do not aim to be exhaustive. Our risk models combine categories of threat actors (ranging from non-technical novice to well-resourced nation states) with harm vectors (such as discovering and exploiting vulnerabilities, spear phishing campaigns, malware development, etc.) and targets (ranging from hardened critical national infrastructure to soft targets). Developers should conduct a risk analysis to comprehensively identify unacceptable risk models, and evaluate at least the `easiest' models, covering the least sophisticated capabilities that may be used to cause harm. Consequently, if the AI system is incapable of enabling these basic scenarios, it is unlikely that there is another area of high concern. Developers should also justify their selection of risk models (\textbf{C3.x}). Justification for the chosen risk models may include reference to risk modelling work done by authoritative third parties or external validation of the risk modelling work done by the developer (\textbf{E3.1}). A challenge for identifying risk models is that methods of attack might not always be publicly disclosed because of national security concerns (\textbf{D3.1}). Collaborating with the government, in particular national cybersecurity authorities, may mitigate this failure mode.

\subsection{Risk models → Proxy tasks}\label{risk-models-proxy-tasks}

We now discuss the second part of the safety case template on a node-per-node basis.

\begin{figure}[t!]
    \centering
    \includegraphics[width=\linewidth]{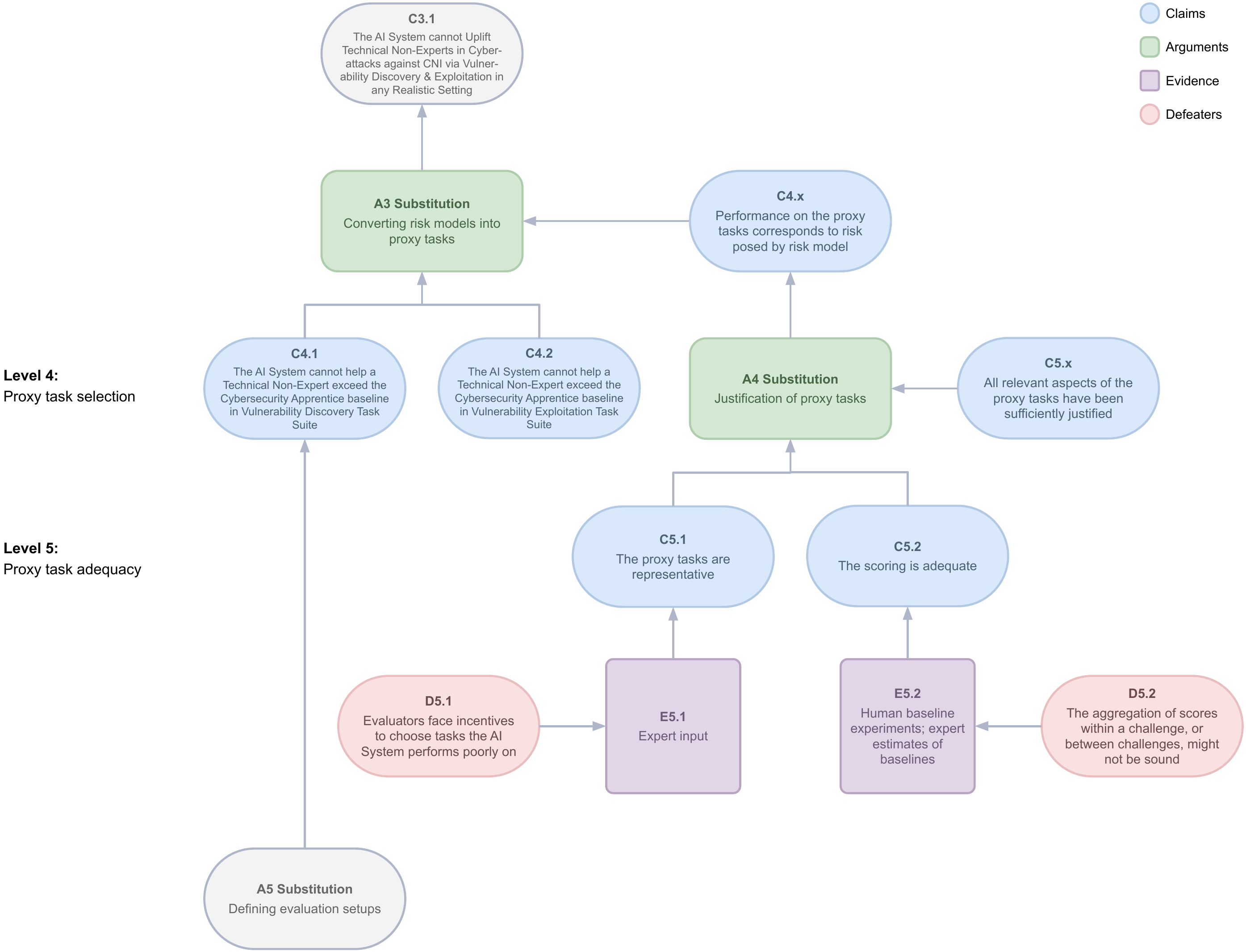}
    \vspace{-0.5em}
    \caption{Part 2 of the safety case template}
    \label{fig:4}
\end{figure}

\textbf{Level 4: Proxy task selection}. Each risk model in \textbf{C3.1} to \textbf{C3.N} specifies a high-level capability threshold that the AI system cannot cross. The safety case must now explain how the model was evaluated on each capability. This safety case template focuses on the risk model identified in C3.1. It follows a commonly used method for evaluating dangerous capabilities, using proxy tasks to assess model performance (\textbf{A3}). Proxy tasks are concrete tasks that are feasible to test the model on, and for which performance is correlated with performance on the high-level capability. \textbf{C4.1} and \textbf{C4.2} specify the proxy tasks and set performance thresholds. The proxy tasks chosen for this example are a vulnerability discovery task suite and a vulnerability exploitation task suite, each consisting of Capture-The-Flag (CTF) challenges\footnote{CTFs are exercises designed to assess cybersecurity skills---solving them requires leveraging cyber expertise to find a `flag', which is usually a random string that is hidden in the target system. CTFs are flexible in that they can be custom-made to test a variety of cyber skills at various levels of difficulty. Even complex multi-step cyber range tasks can usually be converted into this format. In addition to being flexible, CTFs are easily gradable. Either the flag is captured or it is not, which allows for fast automated scoring.}. CTF challenges are easily gradable and can be designed to test a variety of cyber skills at various levels of difficulty. The discovery suite could draw from pre-existing evaluation frameworks for CTF challenges, such as Cybench \cite{zhang2024a}. Specific tasks may include identifying which file contains a vulnerability, and identifying the vulnerable code in such a file \cite{zhang2024a}. For each task suite, there are performance thresholds corresponding to the capability levels of various threat actors. The claim `cannot uplift' in C3.1 means that the AI system cannot improve the performance of the Technical Non-Expert using the AI system to a capability level of a higher-tier threat actor (Cybersecurity Apprentice or above).

\textbf{C4.x} asserts that the proxy task selection and performance thresholds are adequate. The proxy tasks are adequate if the AI system's failure to perform above the threshold in either task suite implies that the risk model in C3.1 is of no concern. This means that the proxy tasks should accurately reflect real-world conditions. This is assessed in the next step.

\textbf{Level 5: Adequacy of proxy tasks.} \textbf{A4} breaks down the evaluation setting into its components to demonstrate the adequacy of each component.

\textbf{C5.1} addresses the adequacy of the proxy tasks. For example, the CTF challenges included in the vulnerability discovery task suite are adequate proxy tasks if they serve as strong indicators of the AI system's ability to help a Technical Non-Expert identify vulnerabilities in CNI in a realistic setting. This could be the case because the tasks required to pass the selected CTF challenges are similar to the necessary steps in this risk model, or because they otherwise serve as strong predictors of success in tasks that are essential for vulnerability discovery.

\textbf{C5.2} addresses the adequacy of the proxy task \emph{threshold}, which functions as the scoring system for these tasks. This scoring system is adequate if it distinguishes performance levels that correspond to the AI system's utility for various actors in real-world scenarios. This requires accurate baselining of the unaided performance of different threat actors. For instance, the evaluators should have an understanding of how a Technical Non-Expert and Cybersecurity Apprentice relying solely on internet search might perform on the task suite, in order to estimate the scores that translate into practical benefit for real-world vulnerability discovery. The ideal method to establish such baselines is through experiments, which calibrate the scoring system according to the observed performance of real-world actors. However, these experiments face limitations in ecological validity, as the motivations and practical constraints of participants in a study may not adequately represent real-world settings.

\textbf{E5.1} and \textbf{E5.2} note that the proxy task adequacy can be substantiated by experiments (as explained above) as well as expert input---where experts estimate how proficient a given threat actor would be at the task suite. The representativeness of proxy tasks and thresholds could also be supported by prediction evals, which assess an actor\textquotesingle s ability to accurately forecast how model evaluation outcomes will generalise to different, untested conditions \cite{chan2024a}.

\textbf{D5.1} and \textbf{D5.2} underscore several challenges with this approach. First, evaluators could face incentives to choose the tasks the AI system performs poorly on, possibly due to commercial pressures to move forward with deployment. Second, because of interaction effects, the aggregation of scores within or between task suites may not represent the full breadth of real-world capability. External validation of the proxy task selection or adherence to a task standard \cite{metr2024} can provide supporting evidence that the selected proxy tasks and thresholds are adequate.

\textbf{C5.x} claims that all relevant aspects of the proxy tasks have been identified, such that if the assessed aspects (i.e. the task selection and the thresholds) are adequate, then the proxy tasks are adequate overall. This is left as an assumption unsupported by evidence.

\subsection{Proxy tasks → Evaluation}\label{proxy-tasks-evaluation}

We now discuss the third and final part of the safety case template on a node-per-node basis.

\begin{figure}[t!]
    \centering
    \includegraphics[width=\linewidth]{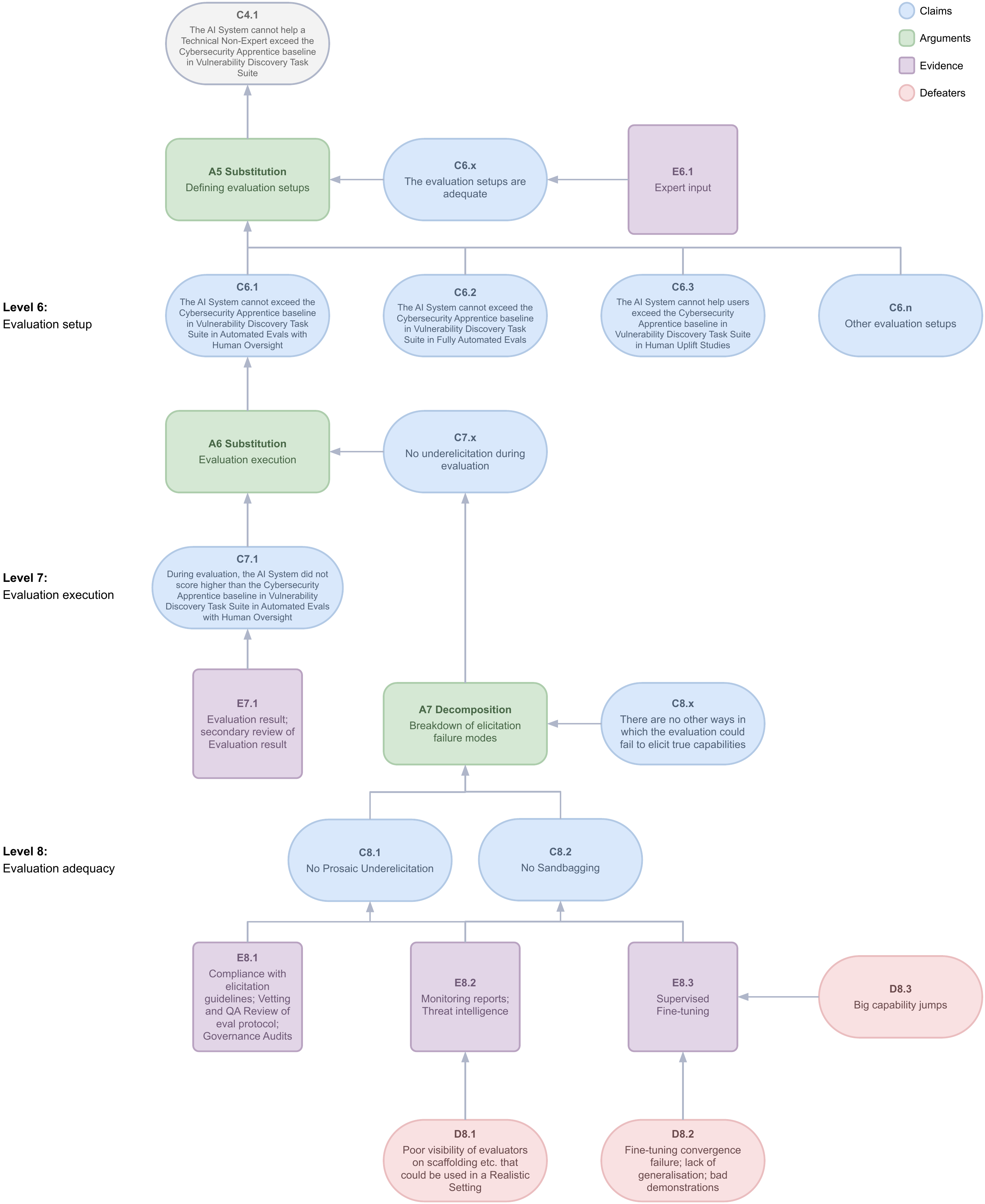}
    \vspace{-0.5em}
    \caption{Part 3 of the safety case template}
    \label{fig:5}
\end{figure}

\textbf{Level 6: Evaluation setup.} There are different ways in which performance on the proxy tasks can be measured, for example through fully automated evaluations, automated evaluations with human oversight, or human uplift experiments. In a human uplift experiment, humans interact with the AI system through a conversational interface to complete tasks. By contrast, automated evaluations require the AI agent to solve tasks, either independently or with some hinting or oversight by an evaluator. Human uplift studies offer empirical insight into how an AI system could lower the skill floor for threat actors to cause harm, though they are expensive and methodologically challenging (see discussion under C5.2). Automated evaluations, on the other hand, are important to cover indirect paths to harm. A threat actor with technical skills could create an agent powered by the AI system (e.g. using tools like LangChain \cite{langchaininc.2022}). This agent could be more capable than the threat actor directly interacting with the AI system, thus warranting separate evaluation of this scenario. Depending on the risk model, a number of evaluation setups are defined (\textbf{C6.1} to \textbf{C6.n}). To ensure the selection adequately addresses the risk model (\textbf{C6.x}), the choice of evaluation setups should be substantiated with expert input (\textbf{E6.1}), which may involve documenting existing practices, analysing assumptions about the evidentiary value of each setup, and seeking external validation.

\textbf{Level 7: Evaluation execution.} \textbf{A6} converts the claim about the AI system's capabilities into a specific claim about the evaluation result in a given evaluation setup (\textbf{C7.1}). This claim is substantiated by the specific results of the evaluation. Evaluation results may be supplemented by further evidence to ensure the reliability of these results, such as independent reviews or secondary analyses (\textbf{E7.1}). This is intended to rule out material errors in the evaluation execution or in the interpretation of the results. Moreover, \textbf{C7.x} assures that there was no underelicitation of capabilities. This claim ensures that the evaluation process was not only free from errors but also elicited the model's capabilities to the fullest extent required by the risk model.

\textbf{Level 7: Evaluation adequacy.} The assessment of evaluation adequacy depends on the evaluation setup used. For instance, in the context of a human uplift experiment, adequacy would depend on factors such as the representativeness of the sample (e.g. whether the study participants possess skill levels similar to the real-world threat actors they are meant to represent) and incentives of the study participants (e.g. whether their level of determination mirrors that of a real-world threat actor).

In our template, however, we provide an example of an adequacy assessment for an automated evaluation with human oversight, focusing on the technical aspects of the elicitation process. \textbf{A7} breaks down underelicitation into two possible failure modes: prosaic underelicitation (\textbf{C8.1}) and sandbagging (\textbf{C8.2}). Both can create a discrepancy between the system's actual capabilities and the evaluation result.

In cases of prosaic underelicitation (\textbf{C7.1}), evaluators do not exert adequate rigour to elicit the AI system\textquotesingle s capabilities. What is adequate depends on the risk model that is being assessed. Risk models involving less sophisticated threat actors may be sufficiently evaluated with limited elicitation efforts, whereas risk models involving more sophisticated threat actors require a more thorough elicitation process. The most demanding standard is state-of-the-art (SOTA) elicitation. Besides prompt engineering, SOTA elicitation attempts will typically involve enhancing the model with tools (such as a web browser) and agentic scaffoldings as well as fine-tuning the model on specific datasets---all of which can dramatically improve a model's performance on a task \cite{davidson2023,openai2024,shinn2023,yang2024,yao2023}. To ensure SOTA elicitation, the search for optimal prompts and enhancements should be thorough. Evaluators should conduct multiple rounds of iteration of prompts, fine-tuning and scaffolding against the development set of an evaluation, continuing this process until no spurious failures are observed and the efforts begin to yield diminishing returns \cite{metr2024a}. Ideally, elicitation efforts should not cease unless there are no known approaches that will predictably yield large performance improvements.

To ensure that the evaluators stay on the frontier of the most successful elicitation techniques, the evaluator could continuously monitor leaderboards of LLM agent benchmarks such as GAIA \cite{mialon2023}, AgentBench \cite{liu2023}, SWE-bench \cite{jimenez2024}, MLE-bench \cite{chan2024} or Cybench \cite{zhang2024a} as well as other channels (including developer and technical communities like GitHub or preprint repositories like ArXiv) that may offer visibility on what enhancements are being applied on similar, deployed models (\textbf{E8.2}). Such visibility is always imperfect, and some actors have incentives to refrain from disclosing improvements (\textbf{D8.1}). However, sufficiently conservative capability thresholds can serve as a mitigation measure, ensuring that an uptick in performance in the real world does not necessarily translate into problematic capabilities.

Regardless of the elicitation standard required, the elicitation process should be documented and the report, together with transcripts of individual runs, should be subjected to independent review. Besides this documentation, more indirect evidence of adequate elicitation efforts consists of adherence to external elicitation guidelines (see e.g. the METR guidelines \cite{metr2024a}) and the result of governance audits (including ISO 42001 audits) that assess internal quality management processes (\textbf{E8.1}). In particular, evidence of supervised fine-tuning can corroborate the adequacy of the elicitation process (\textbf{E8.3}). However, there are limits to the reliability of this approach. It could be the case that more sophisticated threat actors, in particular nation states, have high-quality data sets (for example, a tailor-made, non-public database of vulnerabilities and exposures) that the evaluators do not have access to. This undermines the evidentiary value of fine-tuning efforts by the evaluator, as better demonstrations can result in better performance (\textbf{D8.2}). Again, sufficiently conservative capability thresholds can serve as a mitigation measure.

Sandbagging (\textbf{C8.2}) is a hypothetical second failure mode in the elicitation process. It occurs when an AI system that is being evaluated deliberately underperforms for strategic purposes, such as avoiding shutdown \cite{weij2024}. While current frontier models are very unlikely to be capable of such deceptive behaviour, some authors argue it is a plausible consequence of increasing sophistication of goal-directed AI systems \cite{carlsmith2023,hubinger2024,ngo2024,weij2024}. Supervised fine-tuning can provide assurances that an AI system is not obfuscating its capabilities (\textbf{E8.3}), since a scheming model would not be able to bypass fine-tuning elicitation efforts without a sudden and significant jump in capabilities \cite{greenblatt2024} (\textbf{D8.3}). Monitoring the scaling trend of a given model family\footnote{In particular, monitoring how pretraining loss or downstream task performance changes as a function of model size or compute.} may in turn offer some evidence against such abrupt capability jumps \cite{hoffmann2022,hu2024,pearce2024,sardana2024,schaeffer2023}. However, even absent significant shifts in capabilities, fine-tuning could still fail to elicit latent capabilities (\textbf{D8.2}). This could happen if the demonstrations of successful task completion provided during fine-tuning are suboptimal for a given model.\footnote{I.e. the model cannot learn to complete the task following the demonstrations but is able to come up with another strategy that is successful and that the evaluators did not think of.} Alternatively, fine-tuning could fail to converge and generalise to unseen demonstrations. While we do not elaborate on these defeaters in this paper, a comprehensive safety case should stipulate the mitigations in place to address each of these potential challenges. Overall, sandbagging is less relevant in a safety case that strictly addresses misuse concerns. In the future, we expect that inability arguments could be integrated into a holistic safety case that also evaluates alignment.

Finally, \textbf{C8.x} claims that, with prosaic underelicitation and sandbagging, all key sources of underelicitation have been identified. This is left as an assumption unsupported by evidence.

\section{Conclusion}\label{conclusion}

In this article, we conceptualise a safety case template for offensive cyber capabilities. We suggest how a high-level claim about the acceptability of the cyber risk associated with a given AI system can be broken down into progressively specific sub-claims. We give examples of the types of evidence that could be used to support sub-claims, as well as potential defeaters that could undermine the reliability of such evidence. This method integrates high-level safety claims, risk models, proxy tasks, and evaluation results within a Claims Arguments Evidence framework. While uncertainties around the specifics remain, this template serves as a proof of concept, offering developers and regulators a foundation to guide the development of a more comprehensive safety case methodology.

Based on the findings of this article, we suggest the following questions for further research. First, more work is needed to develop granular evaluations that correspond to specific risk models. In particular, it would be helpful to address the limitations of human uplift experiments. The observed performance of real-world actors should serve as the standard for capability baselining of various threat actors, but these experiments face limitations in ecological validity, as the motivations and practical constraints of study participants may not adequately represent real-world settings. A lack of reliable human uplift data reduces the credibility of an inability safety case.

Second, while we highlight a number of defeaters, further research is needed to develop a more systematic approach to addressing defeaters in frontier AI safety cases. Such an approach would ensure that necessary safety mitigations are comprehensively identified, leading to a more integrated and robust assurance framework. Besides specifying mitigations, a more systematic approach to addressing defeaters could also include establishing their relative importance. This would help identify the key weaknesses in the safety argument and underscore which mitigations are most critical.

Third, developing a more nuanced framework for human uplift is essential. This should account not only for AI systems enhancing the skill level of threat actors, but also for ways AI might enable larger-scale or more frequent attacks. Further research is needed to create a framework that explores the various ways AI can increase both the sophistication and volume of cyber threats. Relatedly, further research could consider how safety cases might need to address potential threats from AI systems acting autonomously rather than merely assisting human threat actors. While such capabilities remain speculative at this stage, expanding the framework beyond human uplift could prove valuable if these concerns become more pressing.

Fourth, our claims are binary---it is implied that, if substantiated, a claim is 100\% true. In reality, we expect that developers would have varying levels of confidence in claims of this nature. Assigning probabilities to each claim and propagating this throughout the structure might improve the accuracy of the safety case and provide a more measured perspective on the degree of assurance that the safety case offers. This could, in turn, enable discussions about the degree of confidence that is required in development and deployment decisions for frontier AI, as compared to other safety critical industries.

Fifth, the opacity of frontier AI systems in combination with their open-endedness and large action space make it challenging to assert the completeness of any safety analysis. Often, crucial questions remain open: \emph{Have we identified all relevant risk models? How many evaluation setups are enough?} Dealing with this uncertainty is perhaps the core challenge for frontier AI safety cases. Future research could explore coping strategies, for example by exploring how different sources of inadequate evidence can be synthesised into adequate evidence, or how safety assessments can be updated as new evidence becomes available.

We believe developing safety case templates for various AI risk domains and safety arguments is essential to foster structured discussions on AI safety. We hope this work provides a foundation for such discussions, thereby advancing the frontier of AI assurance.

\section*{Acknowledgements}\label{acknowledgements}

We would like to express our gratitude to the people who have offered feedback and input on the ideas in this paper: Sandhini Agarwal, Markus Anderljung, Robin Bloomfield, Nate Burnikell, Michael Chen, Nicola Ding, Ben Garfinkel, Friederike Grosse-Holz, John Halstead, Marius Hobbhahn, Leonie Koessler, Sébastien Krier, Anne le Roux, Jade Leung, Vlad Mikulik, Nikhil Mulani, Caitlin O'Kelly, Mary Phuong, Luca Righetti, Gaurav Sett, Rohin Shah, Tony Wang, and Peter Zhang.


\bibliographystyle{abbrv}
\bibliography{ms}

\clearpage

\appendix

\section*{Appendix A: Definitions}\label{appendix-a-definitions}

{\renewcommand{\arraystretch}{1.5}
\begin{longtable}{@{}>{\raggedright\arraybackslash}p{0.23\linewidth}>{\justifying\arraybackslash}p{0.72\linewidth}@{}}
\toprule
\textbf{Term} & \textbf{Definition} \\ \midrule
\endfirsthead
\toprule \textbf{Term} & \textbf{Definition} \\ \midrule
\endhead
\midrule
\multicolumn{2}{r}{\textit{Continues on the next page}} \\
\endfoot
\endlastfoot
AI System &
  The hypothetical AI system that is being assessed in this safety case template. \\
Automated Evals with Human Oversight &
  Assessments where an automated evaluation (tasks with an automated grading mechanism) is run with a   human-in-the-loop. The human has the facility to assist the model or agent by   providing additional prompting, making adjustments to the agent scaffold, or providing other guidance. The level of human intervention can vary to   identify the minimum amount of external support required for the model to   succeed. \\
Cyberattack &
  Any kind of malicious activity that attempts to collect, disrupt,   deny, degrade, or destroy information system resources or the information   itself \cite{nist}. \\
Cyber Risk &
  The potential of injury or damage to the health, or   damage to property or the environment, or interference with the fundamental   rights of a person or organisation as a result of a Cyberattack. \\
Cybersecurity Apprentice &
  An individual with 1-3 years of   cybersecurity professional experience, less-skilled hackers for hire, or   opportunistic cyber criminals. They operate with limited resources, with a   budget of up to \$1,000 and several weeks on the operation, accessing commercial   AI models and basic AI-enabled tools. \\
Cybersecurity Practitioner &
  An individual with 3-10 years of   professional cybersecurity experience, specialising in some technical domain   of the field, e.g. software developers or Security Operations Centre   analysts. They have a limited budget of up to \$1,000, access to commercial AI   models, and can repurpose AI-enabled commercial penetration testing tools. \\
Critical National Infrastructure (CNI) &
  National assets that are essential for   the functioning of society, such as those associated with energy supply,   water supply, transportation, health, and telecommunications, or a private   asset with similar cybersecurity standards. \cite{ncsc2024} \\
Conventional Cyberattack &
  A Cyberattack that utilises techniques   that are well-understood and currently observed. \\
Fully Automated Evals &
  Assessments   where model performance on tasks is evaluated in an automated fashion.   Examples include Q\&A benchmarks with automated graders or, as is common   in cybersecurity, tasks in a capture-the-flag (CTF) format that provide a   binary grading of success at a task carried out by a scaffolded model. \\
Governance Audit &
  An audit of the organisation developing   the AI Systems, including assessments of software development and quality   management processes, incentive structures, and the allocation of roles and   responsibilities \cite{mokander2023}. \\
Hardened CNI &
  CNI that is continuously fortified with   updated security measures, including advanced cyber defences, routine audits,   and personnel training to resist various threats and maintain robust   operation. \\
Human Uplift Studies &
  Randomised Control Trials that   assess the uplift (i.e. increase in performance) an AI system might provide to a human threat actor at a   predefined task. \\
Moderately Resourced Nation-State &
  Moderately resourced nation states or   commercial cyber intrusion companies that employ numerous different   individuals with expertise in all realms of the field. They have significant   resources, spending up to \$1-10m and several months per operation, incorporating   advanced AI-enabled tools and services, including proprietary information and   specialised scaffolding and tooling, equivalent to OC4 in Nevo et al. \cite{nevo2024}. \\
Non-Technical Novice &
  A person with no technical, coding, or   security expertise who decides to perform an attack nonetheless, e.g. a   disgruntled worker. They have limited resources, with a budget of up to   \$1,000 and several weeks on the specific operation, accessing commercial AI   models and basic AI-enabled tools, equivalent to OC1 in Nevo et al. \cite{nevo2024}. \\
Novel Cyberattack &
  Potential future Cyberattacks that   leverage speculative AI capabilities for currently unknown methods of attack. \\
Prosaic Underelicitation &
  Elicitation efforts that fall short of   the reasonably foreseeable elicitation capabilities of the threat actor in   the assessed risk model. \\
Proxy Task &
  Concrete tasks that are indicative of the   AI System’s abilities for more general tasks. \\
QA Review &
  Review of the evaluation by a human with   task development experience and the expertise required for the task, who also   implements any fixes or changes necessary \cite{metr2024a}. \\
Realistic Setting &
  The prompts and the set of practical and   typical enhancements (e.g. fine-tuning, scaffolding) and affordances (e.g.   compute budget, internet access) with which the AI System may be utilised,   reflecting real-world conditions and constraints. \\
Safeguard &
  A technical or social mechanism designed   to prevent, mitigate, or manage potential risks and harms associated with the   deployment of the AI System. \\
Sandbagging &
  The AI System strategically underperforms   on an evaluation \cite{weij2024}. \\
Soft Target &
  Any asset that is not CNI. \\
SOTA Elicitation &
  An effort to elicit an AI System’s   capabilities using the methods that are currently considered to yield best   performance, including, for example, the best methods for supervised   finetuning, prompting, and agentic scaffolding. \\
Spear Phishing Campaign &
  A type of Cyberattack that involves using   personalised information to manipulate targets into revealing sensitive   information \cite{hazell2023}. \\
Supervised Fine-tuning &
  Training a pre-trained LLM to do a   specific downstream task using demonstrations of completing the task   successfully. \\
Technical Non-Expert &
  A novice with limited or no knowledge in   the realm of cybersecurity, comparable to high school to university level   expertise but with some technical expertise, e.g. a data analyst or   engineer. They have limited resources, with a total budget of up to \$1,000   and several weeks on the operation, accessing commercial AI models and basic   AI-enabled tools. \\
Threat Actor &
  In order of ascending capability:   Non-Technical Novice, Technical Non-Expert, Cybersecurity Apprentice,   Cybersecurity Practitioner, Cybersecurity Expert, Moderately Well-Resourced   Nation State, Well-Resourced Nation State \\
Uplift &
  Use of the AI System to enhance a Threat   Actor’s ability to conduct Cyberattacks to the equivalent of a higher-tier   Threat Actor. \\
Vetting &
  Review of the evaluation procedure by   someone other than the main evaluation developer \cite{metr2024a}. \\
Vulnerability Discovery &
  Finding weaknesses or flaws in a computer   system, network, application, or protocol that allow unauthorised access,   manipulation of data, disruption of operations, and/or cause other malicious   effects. \\
Vulnerability Discovery Task Suite &
  A series of tasks that measure the AI   System’s proficiency in Vulnerability Discovery. These tasks cover a range of   difficulty levels, from identifying a known vulnerability in the code base   given a brief description (e.g. CVE), through identifying vulnerabilities   that can already be found using existing static analysis tools (e.g.   Metasploit), to identifying complex novel vulnerabilities that surpass   existing vulnerability discovery tools. \\
Vulnerability Exploitation &
  Taking advantage of weaknesses or flaws   in a computer system, network, application, or protocol that allow   unauthorised access, manipulation of data, disruption of operations, and/or   cause other malicious effects. \\
Vulnerability Exploitation Task Suite &
  A series of tasks that measure the AI   System’s proficiency in Vulnerability Exploitation. It measures the initial   proficiency of the AI model in executing vulnerability exploitation   techniques, including web-based and OS-level attacks. \\
Well-Resourced Nation State &
  Well-resourced state actors with   significant expertise in all realms of the field. They operate with very   significant resources, spending up to \$10m-\$1b and years per operation, with   state-level infrastructure. These threat actors develop bespoke advanced   systems to gain a strategic advantage, e.g. through finetuning on highly   curated private or sensitive datasets, significant specialised scaffolding,   or multiple expert monitoring and oversight, equivalent to OC-5 in Nevo et al. \cite{nevo2024}. \\ \bottomrule
  \\[-10pt]
\caption{Definitions of key terms used in our safety case template}\label{tab:1} \\ 
\end{longtable}%
}

\clearpage

\section*{Appendix B: Full diagram}\label{appendix-b-diagram}

\begin{figure}[ht!]
    \centering
    \includegraphics[width=0.55\linewidth]{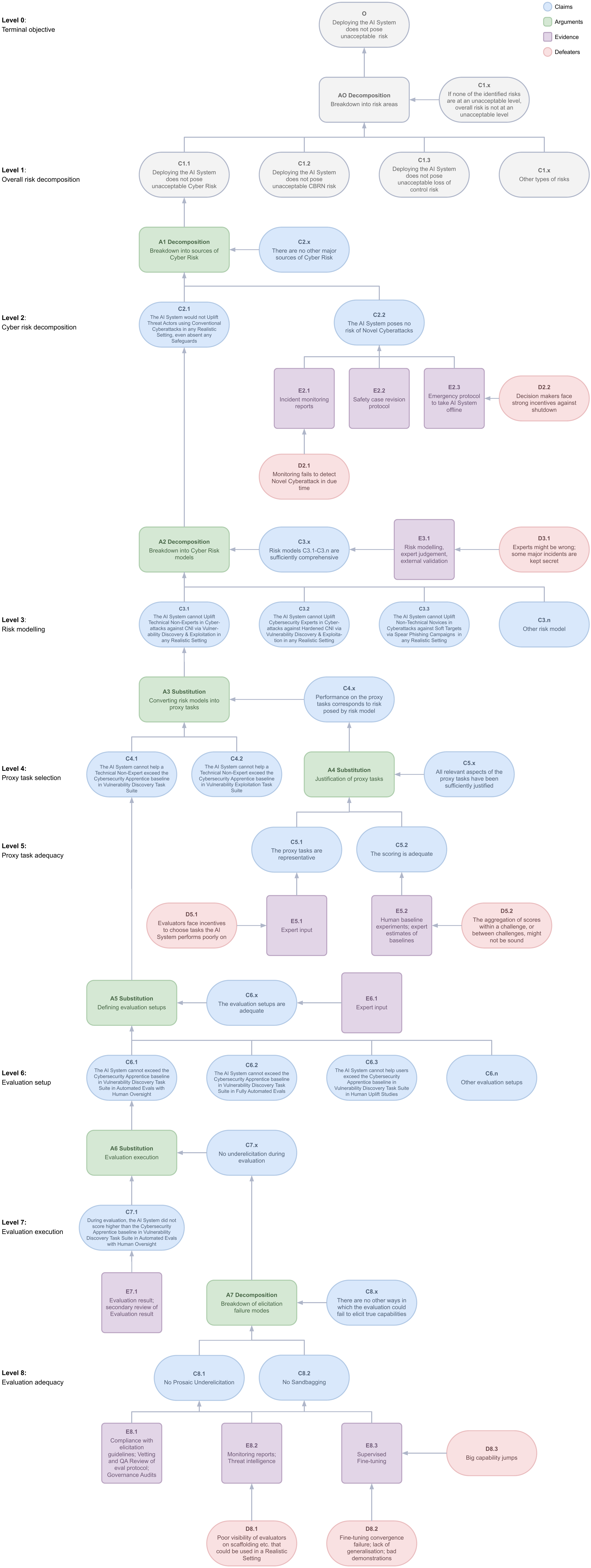}
    \caption{Safety case template}
    \label{fig:6}
\end{figure}

\end{document}